\documentclass[twoside,english,aps,prb,twocolumn,groupedaddress,amsmath,amssymb,floatfix]{revtex4-1}

\usepackage{graphicx}
\usepackage{float}
\usepackage{physics}
\usepackage{amsmath}
\usepackage{dcolumn}
\usepackage{bm}
\usepackage{units}
\usepackage[version=4]{mhchem}
\usepackage{siunitx}
\usepackage{color}
\usepackage{array}
\usepackage{color}
\usepackage{listings}
\usepackage{url}
\usepackage[colorlinks = true, linkcolor = blue, urlcolor  = blue, citecolor = blue, anchorcolor = blue, unicode]{hyperref}

\newcolumntype{P}[1]{>{\centering\arraybackslash}p{#1}}

\def\eV{\,\textrm{eV}}
\def\meV{\,\textrm{meV}}
\def\Ry{\,\textrm{Ry}}

\newcommand{\rom}[1]{\uppercase\expandafter{\romannumeral #1\relax}}

\begin{document}

\title{Insights to negative differential resistance in \texorpdfstring{MoS\textsubscript{2}}{MoS2} Esaki diodes: a first-principles perspective}
\author{Adam V. Bruce}
\author{Shuanglong Liu}
\author{James N. Fry}
\author{Hai-Ping Cheng}
\email{hping@ufl.edu}

\affiliation{Department of Physics, University of Florida, Gainesville, Florida 32611, USA}
\date{\today}

\begin{abstract}
\ce{MoS_2} is a two dimensional material with a band gap depending on the number of layers and tunable by an external electric field. The experimentally observed intralayer band-to-band tunneling and interlayer band-to-band tunneling in this material present an opportunity for new electronic applications in tunnel field effect transistors. However, such a widely accepted concept has never been supported up by theoretical investigations  based on first principles. In this work, using density functional theory, in conjunction with non-equilibrilibrium Green's function techniques and our electric field gating method, enabled by a large-scale computational approach, we study the relation between band alignment and transmission in planar and side-stack \ce{MoS_2} $p$-$i$-$n$ junction configurations.  We demonstrate the presence of negative differential resistance for both in-plane and interlayer current, a staple characteristic of tunnel diode junctions, and analyze the physical origin of such an effect. Electrostatic potentials, the van der Waals barrier, and complex band analysis are also examined for a thorough understanding of Esaki Diodes.

\end{abstract}

\maketitle

\section{Introduction}
Following the advent of the tunnel diode, new efforts emerged to identify candidate materials for such electronics.  
Discovered by Leo Esaki, these devices are valued for their possible applications 
in nanoscale circuitry~\cite{Esaki1958}.
Among these benefits is their usage in tunnel field effect transistors (TEFTs), 
which can have lower theoretical sub-threshold swing than traditional MOSFETs.  
The novel interlayer properties of atomically thin heterostructures have made 
such configurations attractive for TEFT operation.  
The advantage in heterojunction-based tunnel diodes is their difference in band structure between layers.
External bias allows for programmable band alignment between layers, yielding nonlinear characteristics in band alignment.
In particular, graphene and transition-metal dichalcogenides (TMDs) have been 
valued for their suitable dimensionality and electronic structure~\cite{
Britnell2012, YPWang2017, GXChen2017, Radisavljevic2011}.  
One such TMD is \ce{MoS_2}, having a direct band gap in single layer 
configuration~\cite{KFMak2010}.

Recent experiments have shown a highly tunable bandgap in bilayer \ce{MoS_2} 
under external electric field~\cite{TChu2015}.  
Experimentally, there are difficulties in constructing heterojunctions using 2D TMDs as leads~\cite{Kang}.
Semi-empirical studies have observed the presence of band-to-band tunneling in a nanoribbon under strain~\cite{Ghosh}.
In addition, studies finding negative differential resistance in tunnel diode \ce{MoS_2} systems are a fairly recent phenomenon, most being published within the past half-decade~\cite{Nourbakhsh, RZhang}.
Within these studies, there has yet to be a comparison between planar and interlayer transport properties. A first-principles based investigation is necessary to fully understand the electronic processes and the effects of band to band tunneling in $p$-$i$-$n$ configuration \ce{MoS_2} diodes. Before introducing our model for this study, we should mention that we previously have developed and applied first-principles methods to investigate field effects related to tunneling transistors~\cite{Rn2033, RN2183}. A number of systems were studied, focusing on different aspects of physical processes. Some other groups have also performed first-principles investigations to understand field effects in tunneling junctions~\cite{RN158, RN3032, RN162}; however, this study is the first attempt that provides analysis of interlayer band-to-band tunneling in \ce{MoS_2} bilayer systems in addition to a full analysis of band-to-band tunneling in a planar monolayer junction. The two types of junctions are prepared to separately quantify in-plane and interlayer tunneling. The rest of the paper is organized as follows: Section II describes theoretical approaches, computational details, and our simulation models; Section III presents our major results; and Section IV concludes our investigations with discussion of some important issues.

\section{Methods}\label{sec2}

All of our calculations used a first-principles approach. 
Transport calculations were performed using density functional theory~\cite{KS1965, PBE1996}, with non-equilibrium Green's functions ($\mathrm{NEGF}+\mathrm{DFT}$) techniques as implemented in TranSIESTA~\cite{Brandbyge2002, Taylor2001},
which accounts for core electrons via nonlocal, norm-conserving 
pseudopotentials~\cite{Soler_2002, TM1991}.
The self-consistent field (SCF) approach yields the necessary Green's functions to employ the Caroli formula for electron transmission~\cite{Caroli_1971, Meir1992}, 
\begin{equation}
T(E) = \Tr[G^a \Gamma^R G^r \Gamma^L], 
\end{equation}
where $G^a$ and $G^r$ are advanced and retarded Green's functions, respectively, and $\Gamma^R$ and $\Gamma^L$ represent the the level broadening functions for the right and left leads, respectively. This expression for the transmission function allows us to obtain the current across the scattering region via the usual 
Landauer equation~\cite{SLLRN274, Landauer1957}, 
\begin{equation}
I=\frac{e}{h} \int_{\mu_R}^{\mu_L}dE \, [f_L (E)-f_R (E)]T(E),
\end{equation}
where $\mu_L$ and $\mu_R$ are the chemical potentials of the left and right reservoirs, and $f_L$ and $f_R$ are Fermi functions in the left and right leads.
For the basis set, we use a double-zeta plus polarization (DZP) basis.  
We also employ the Monkhorst-Pack method for sampling $k$-points in the 
Brillouin zone~\cite{Monkhorst1976}.
Our rectangular supercell uses a $1 \times 11 \times 1$ grid for SCF calculations and 
a $1 \times 101 \times 1$ grid for all transmission calculations.  
We set the convergence requirement for the probability density matrix to $1 \times 10^{-3}$.
Our functional for the exchange-correlation energy is within the generalized gradient approximation (GGA), specifically, the Perdew–Burke-Ernzerhof (PBE) form~\cite{PBE1996}.
The applicability of this functional to our systems will be discussed in a later section.

In the experiments led by Chen, the metal leads were made by gating \ce{MoS_2} layers either positively or negatively~\cite{TChu2015}.
In our simulation, we used doped leads, which are achieved by the virtual crystal approximation (VCA) as implemented in SIESTA.
The VCA is a good approximation when the atomic number of the dopant atom is close to the atom it replaces, the doping concentration is relatively small, and wavefunctions spread over a large spatial region~\cite{Elliott1975}.
We use neighboring atoms on the periodic table in our doping scheme to comply with the former restriction and choose the two elements Nb and Tc to mimic hole-doped and electron-doped \ce{MoS_2}, respectively. 
The very nature of the transport problem satisfies the latter.  
We apply this doping scheme to linearly shift the band structure of our leads, achieving the necessary non-equilibrium boundary conditions. 
However, the VCA cannot capture all the physics of random doping, which includes disordered scattering from random defects.
Previous calculations show these effects on spin transport in nanojunctions~\cite{PhysRevB.97.014404}.
Due to the non-linear changes to conductance arising from these effects, it is unclear to what degree the VCA causes an overestimation or underestimation of transmission at this particular doping concentration compared to random doping.
For the purposes of our study, we use the VCA to model gated leads rather than randomly doped leads.

We calculated the complex band structure of pristine monolayer and bilayer \ce{MoS2} using the PWCOND part~\cite{SLLRN289,SLLRN290} of the Quantum ESPRESSO package~\cite{SLLRN172, SLLRN288}. 
During self-consistent calculations with Quantum ESPRESSO, we utilized the PBE exchange correlation energy functional~\cite{PBE1996} together with projector augmented-wave (PAW) pseudopotentials~\cite{SLLRN79}. 
The pseudopotentials were generated using the PSlibrary package version 1.0.0.~\cite{SLLRN291,SLLRN292} 
We applied $500$ and $80 \Ry$ energy cutoffs for expanding the charge density and the wavefunctions respectively. 
The energy tolerance for self-consistency was set to $1 \times 10^{-8} \Ry$. 
During complex band calculations with PWCOND, which is based on the self-consistent field previously found in Quantum ESPRESSO, we adopted an energy cutoff of $80 \Ry$ for expanding the wavefunctions in two-dimensional planes as well. 
The number of subslabs was set to $30$ to ensure high accuracy.

\onecolumngrid

\begin{figure}[htb!]
\centering
\includegraphics[width=0.95\textwidth,height=\textheight, keepaspectratio]{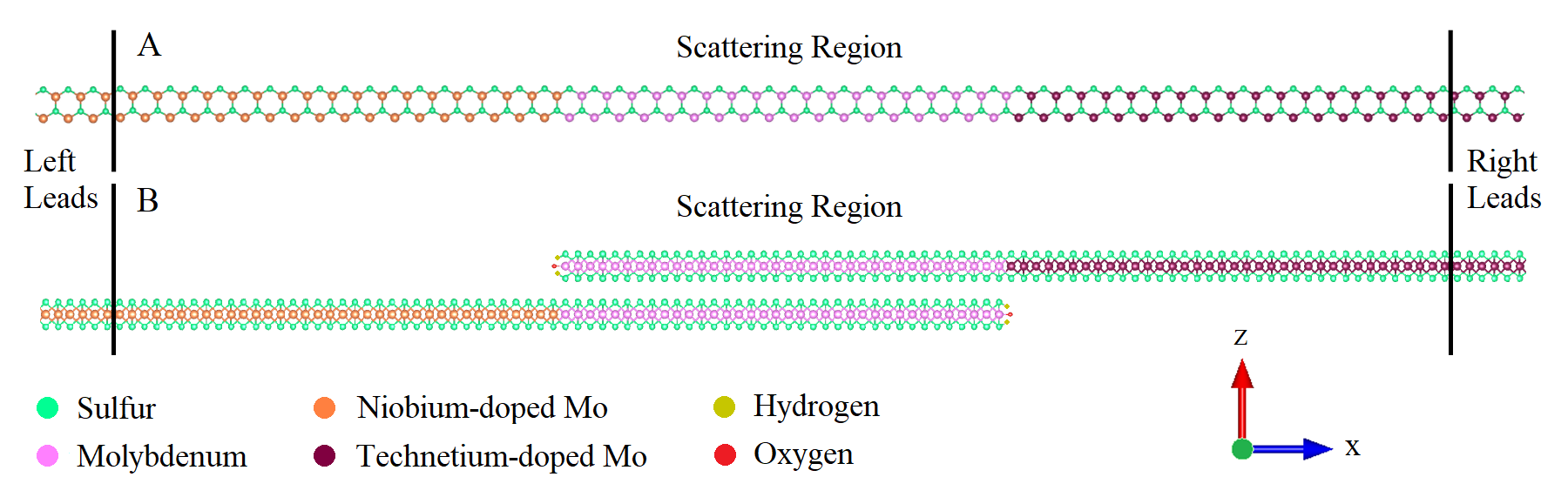}
\caption{\label{Figure 1.} Atomic structure of two molybdenum disulfide junctions (with H- and O-terminated edges). Panel~A is the top view of a $p$-$i$-$n$ junction of single layer \ce{MoS_2} consisting of 10\% niobium and technetium doped leads; Panel~B is the side view of a $p$-$i$-$n$ junction of overlapping, semi-infinite layers of \ce{MoS_2} with overlap in the undoped region. The vertical black lines mark the boundary of the scattering regions. For panel B, the left electrode is connected to the lower layer while right electrode is connected to the upper layer. Atomic positions are drawn in the \emph{VESTA} visualization software~\cite{Momma:db5098}.}
\end{figure}


\twocolumngrid

\section{Results}\label{sec3}

\begin{figure}[htb!]
\centering
\includegraphics[width=0.45\textwidth,height=\textheight, keepaspectratio]{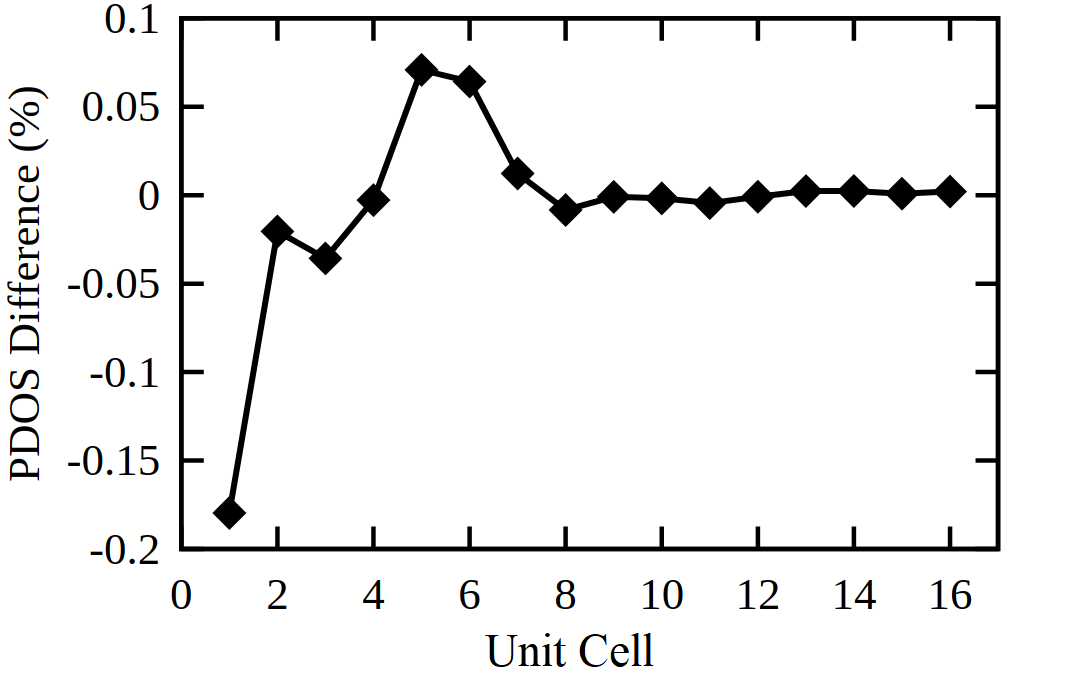}
\caption{\label{Figure DPDOS}Percent difference of the electrode and scattering region partial density of states.  The difference is a function of number of layers of unit cells from the pristine contact.  This result shows the length of the doped contact in the scattering region necessary to approach bulk-like PDOS.
}
\end{figure}

Figure \ref{Figure 1.}A shows our monolayer structure that consists of a central pristine armchair \ce{MoS_2} ribbon  of 18 unit cells in width (or length in the transport direction), bracketed by 18 niobium- ($10\%$) or technetium-doped ($10\%$) \ce{MoS_2} cells on each side and then by the two leads. The 18 unit cells right next to the pristine region or the scattering barrier serve as a buffer zone that screens charge accumulated next to the pristine region such that the whole scattering region is charge neutral. The pristine \ce{MoS_2} ribbon has a finite band gap thus functioning as the tunneling barrier.   
The left and right leads are also \ce{MoS_2} doped with $10\%$ Nb and Tc respectively to create a $p$-$i$-$n$ junction configuration.  
For this doping concentration, we performed a test for the screening length necessary for the transport calculation; we found that the size of the buffer zone is of adequate length to satisfy the charge neutrality condition and that the hopping parameters between the leftmost/rightmost unit cell in the buffer zone to the first unit cell of left/right lead is equal to the hopping between two adjacent unit cells in the left/right lead.

We analyzed the partial density of states (PDOS) according to 
\begin{equation}
\textrm{PDOS} = \frac{1}{\pi} \int d\textbf{k}\, G^r[\Gamma_L + \Gamma_R]G^a,
\end{equation}
in which the Green's function and level broadening functions are the same as those of the Caroli forumula.  A detailed comparison of the PDOS of our buffer zone or depletion region and electrode is illustrated in figure \ref{Figure DPDOS}. Such a comparison is necessary considering the need for exact matching of electronic density in the electrode and scattering region; this test indirectly verifies this agreement.  It is clear that beyond 9 unit cells, the difference in the projected density of states between the buffer zone and a perfect lead is relatively small.

Figure \ref{Figure 1.}B shows the configuration of our second junction. Compared to the planar configuration in Figure \ref{Figure 1.}A, there is an overlapping region in the $z$-direction that introduces a complexity in the junction and also allows interlayer electron tunneling. Again, the undoped insulating \ce{MoS_2} functions as the tunneling barrier.  The configurations of the two junctions allows us to separately quantify in-plane and interlayer transport properties. Previous experimental studies have used similar configurations with heterojunctions and homojunctions 
to study out-of-plane tunneling current~\cite{Agarwal2014, Roy2015}.
Due to the sheared edge of the interlayer junction, hydrogen and oxygen serve as edge-terminating atoms. We will discuss edges and edge termination in a later section.

\begin{figure}[htb!]
\centering
\includegraphics[width=0.46\textwidth,height=\textheight, keepaspectratio]{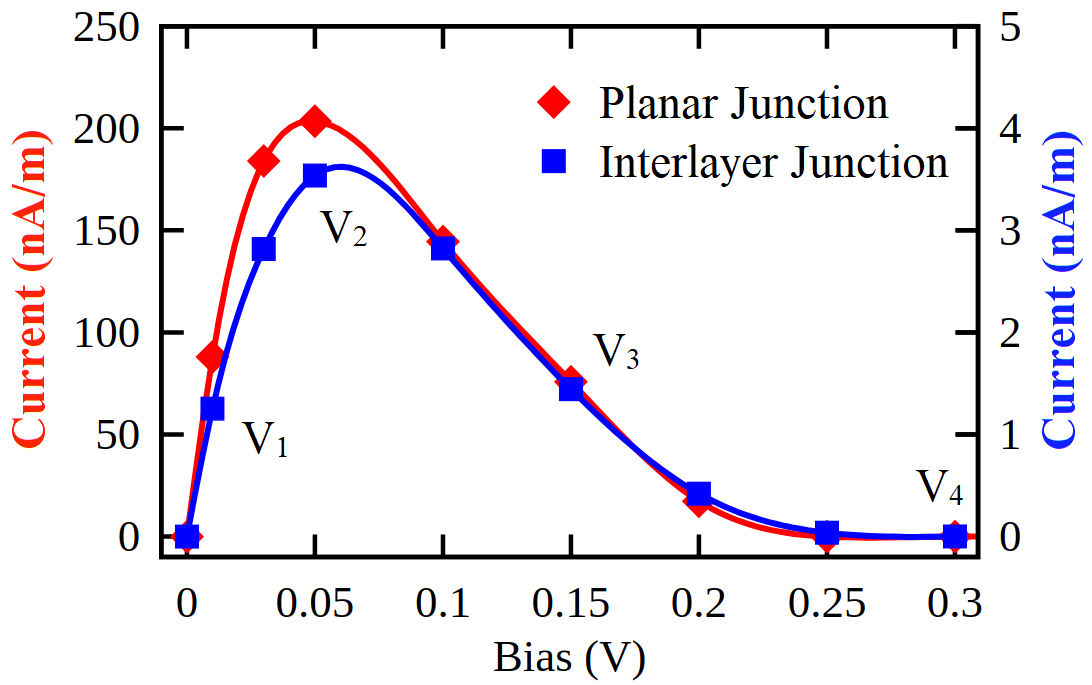}
\caption{\label{Figure 2.}The IV curves for single-layer and overlap \ce{MoS_2} junctions respectively.
The scales for the two curves, left for the planar and right for overlap, differ by a factor of 50.
Both junctions show negative differential resistance.
}
\end{figure}

Figure \ref{Figure 2.} shows the calculated IV curves of both the planar and the side stacking or overlap junctions under low biases at room temperature. 
The characteristics of both graphs are similar, showing negative differential at biases 
higher than approximately $ 0.05 \eV $.  
Above roughly $ 0.25 \eV $, the current of both junctions drops by several orders of 
magnitude relative to the maximum current.  
These IV curves are consistent with that of an Esaki diode in forward bias mode, containing a local maximum at low bias followed by a negative differential resistance regime~\cite{Esaki1958}.
The current in the planar junction reaches a maximum value roughly 50 times that for the junction of the side stacking configuration, indicating the tunneling barrier across the two \ce{MoS_2} layers interacting by van der Waals force is substantially higher than that across the the 18 unit cells of \ce{MoS_2}. 
Using partial density of states analysis, we can evaluate the band character 
of the junctions at biases of interest.  

\begin{figure}[htb!]
\centering
\includegraphics[width=0.46\textwidth,height=\textheight, keepaspectratio]{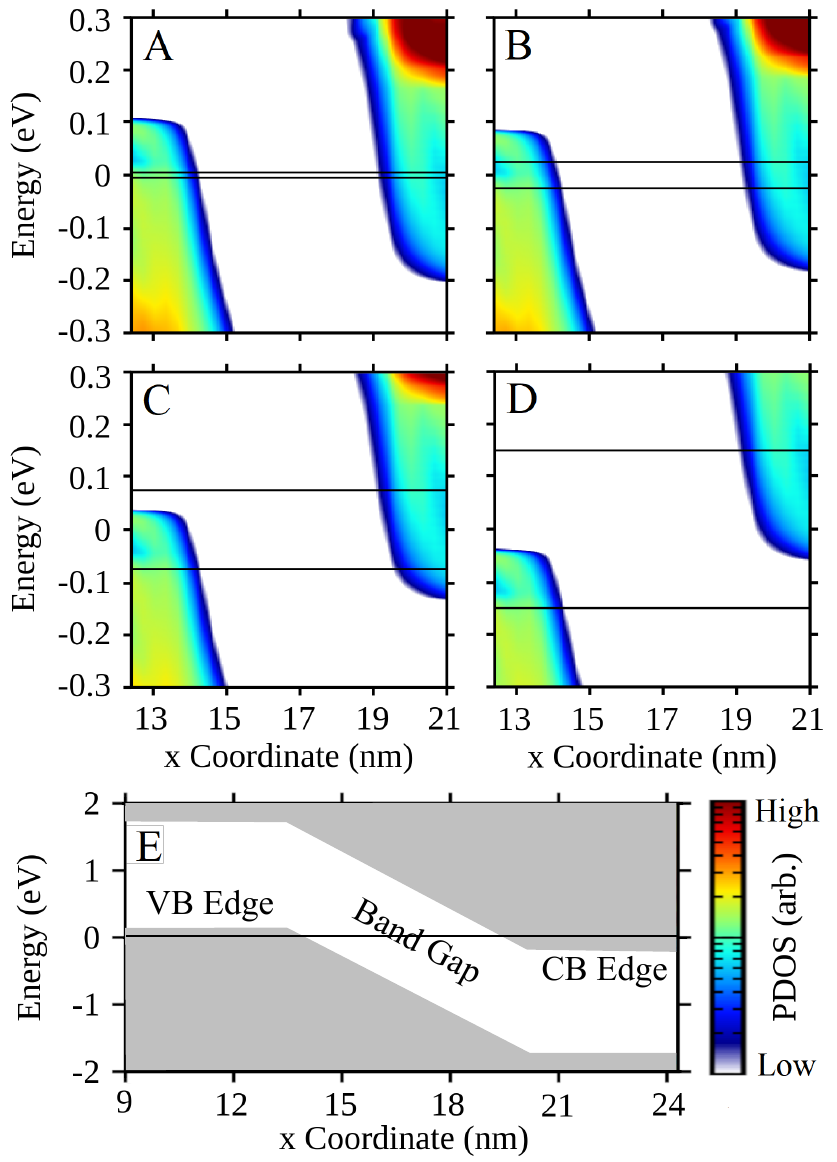}
\caption{\label{Figure 2.5.}A--D: The partial density of states of the planar junction at biases V\textsubscript{1}--V\textsubscript{4} in figure 3 respectively. The color bar shows the magnitude of the PDOS for A--D in logarithmic scale. E: Band diagram at zero bias on a wider energy and position scale. VB and CB represent valence band and conduction band respectively.
}
\end{figure}

\begin{figure}[htb!]
\centering
\includegraphics[width=0.5\textwidth,height=\textheight, keepaspectratio]{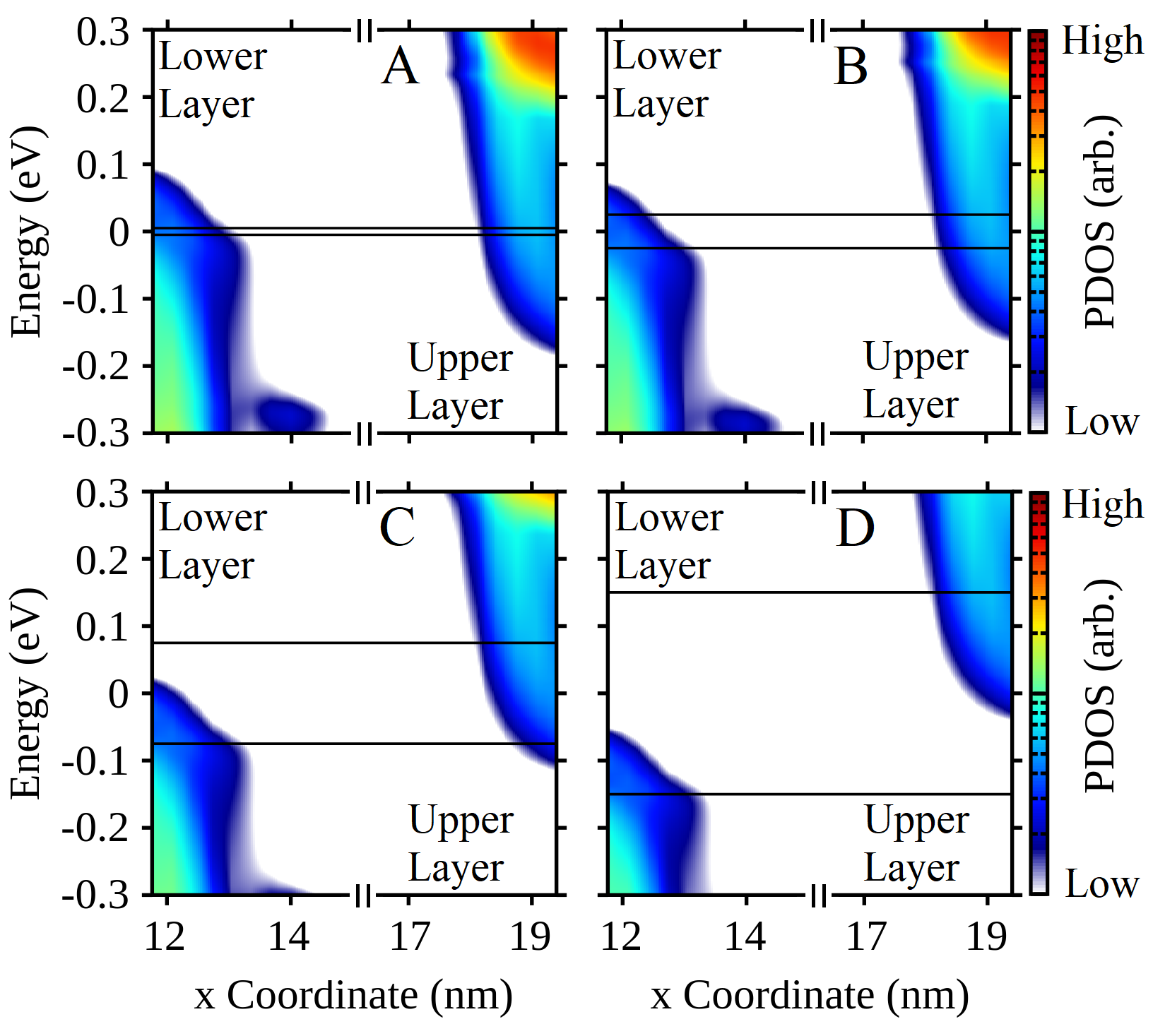}
\caption{\label{Figure 3.}
A--D: The density of states of the interlayer junction corresponding to biases V\textsubscript{1}--V\textsubscript{4} in figure 3 respectively.
The density across the junction is separated into lower and upper layers. 
Band alignment follows the same characteristics as that of the planar junction.
}
\end{figure}

Using the GGA functional, we calculated the band-gap of monolayer \ce{MoS_2} to be 1.59 eV.
This result differs from the measured value of greater than 1.9 eV~\cite{KFMak2010}.
With such discrepancy, we expect an overestimation of tunneling current within our junctions.
However, if the current arises from band to band tunneling effects, we do not expect the size of the band-gap to impact the overall character of our IV curve.
This is due to band to band tunneling being driven by relative band alignment.
Additionally, the measured band-gap of monolayer \ce{Mos_2} can vary as much as 1.39 to 2.16 eV due to environmental dielectric screening~\cite{Ryou2016}.
The discrepancy between measured and our calculated band-gap is within this range of variations caused by environmental dielectric screening, which would be present in many electronic system.
Thus, our calculated band-gap could coincide with the measured band-gap of monolayer \ce{Mos_2} in the presence of certain dielectrics.
For these reasons, we believe that the GGA is sufficient for our calculations.

By plotting the PDOS of our planar junction at the biases of interest in Figure \ref{Figure 2.5.} as a function of position along transport direction and energy, we illuminate the transport properties of our systems.
At V\textsubscript{1}, the valence bands of the left electrode and the conduction bands of the right electrode are well aligned; however, the bias window is relatively small.
The bands remain well aligned at V\textsubscript{2} with bias significantly increased.
Further increases in bias skews band alignment, as seen in V\textsubscript{3} and V\textsubscript{4}, with alignment completely lost at the latter bias.
Band to band tunneling can no longer contribute to current for biases greater than V\textsubscript{4}.
Band alignment within the bias window correlates with higher current, 
characteristic of an Esaki diode.  

\begin{figure}[htb!]
\centering
\includegraphics[width=0.46\textwidth,height=\textheight, keepaspectratio]{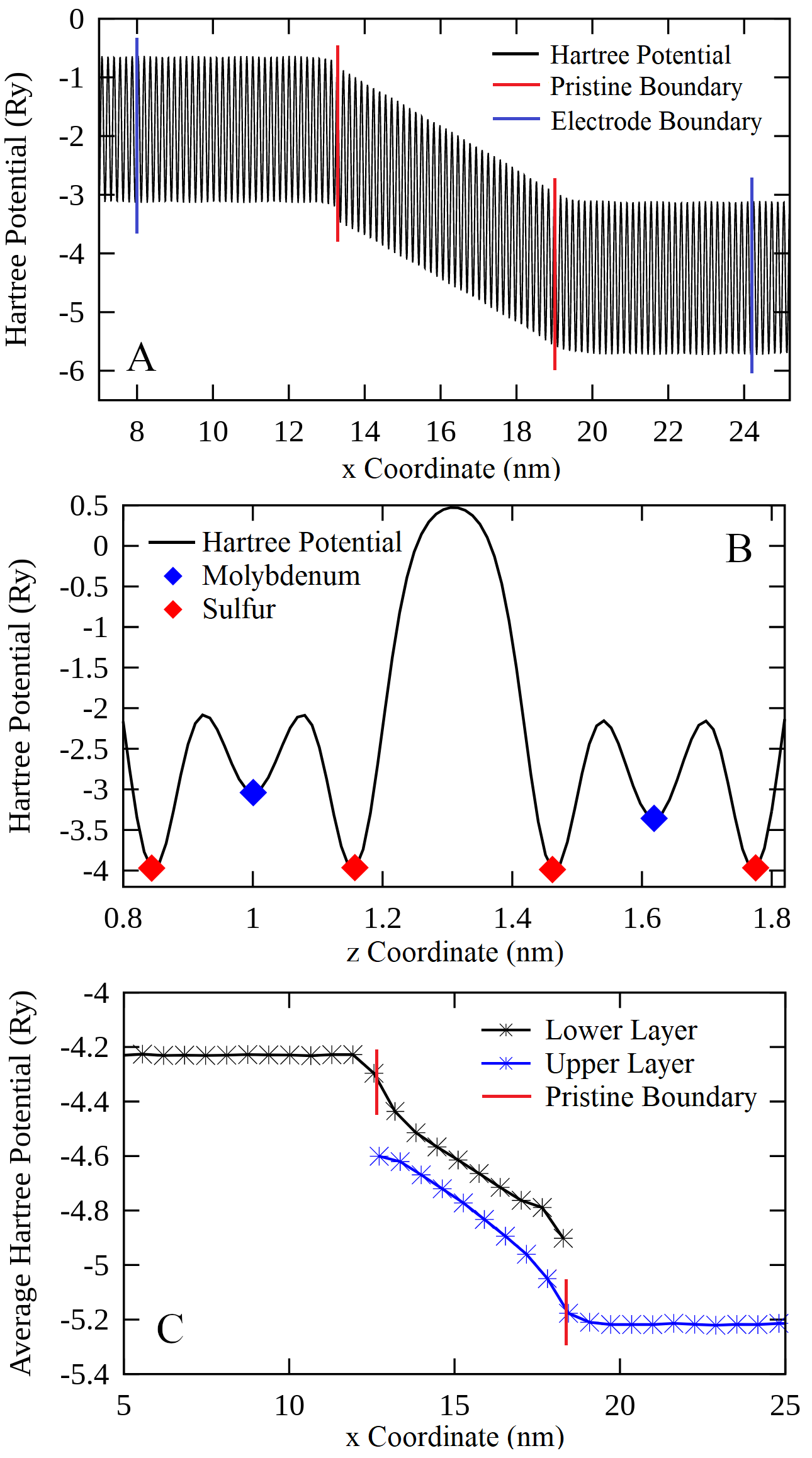}
\caption{\label{Figure H.}The Hartree potential for both junctions. A: The planar averaged Hartree potential of the planar junction as a function of $x$-position. B: The planar averaged Hartree potential of the interlayer junction as a function of $z$-position. C: The average Hartree potential for one unit cell of MoS\textsubscript{2} in the interlayer junction as a function of $x$-position.
}
\end{figure}

In Figure \ref{Figure 3.}, we illustrate the PDOS as function of position and energy of the interlayer junction at the biases of interest.
Comparison between PDOS and current indirectly shows intralayer and interlayer band to band tunneling across our junctions. 
Both planar and interlayer junctions share a similar character in density of states. However, the tunneling barriers are quite different.
Figure \ref{Figure H.} demonstrates this difference via the differing character of the Hartree potential between the junctions. The Hartree potential of the planar junction, seen in figure \ref{Figure H.}A, can be plotted along the transport direction without discontinuity. As illustrated in Figures~\ref{Figure H.}B-C, the interlayer junction has a potential barrier corresponding to the van der Waals gap. Figure~\ref{Figure H.}C shows the relative difference in Hartree potential between the two layers.

If band-to-band tunneling is the primary mechanism of transport across our junctions, we expect that transmission will coincide with the alignment of electrode bands.
Comparison of both electrode band structure and the transmission function is seen in figure \ref{Figure 4.}A-B.  
We see in the band structure of our electrodes an overlap around the Fermi energy; 
the broadest energy range of this overlap is at the $\Gamma$ point in the first Brillouin zone. 
Alongside this band structure is the transmission function of the interlayer junction plotted in a logarithmic scaling.  
We illustrate the correlation of electrode band crossing and transmission function by plotting both functions at the same energy and $k$-path. 
Regions of highest band crossing coincide with regions of highest transmission.

Previous studies show the effects of edge \ce{MoS_2} atoms on electronic structure~\cite{LZhang2015}.
Charge accumulation on edge atoms leads to deviation from bulk electron density.
In order to mitigate these effects, we follow the edge terminating structure of relaxing H and O onto bare \ce{MoS_2}.
We analyze the contribution to local density of states (LDOS) of edge pristine \ce{MoS_2} unit cells via fat bands. 
These are compared to the LDOS of the unit cells adjacent to doped \ce{MoS_2} 
in Figure \ref{Figure 4.}C. 
The edge unit cells show a high contribution to the density of states at the bottom 
of the conduction band. 
As these states fall within our bias window, we must ensure that edge states 
do not connect to the electrodes. 
In Figure \ref{Figure 4.}D, we show the lowest energy, conduction band wavefunction 
of the lower-layer edge unit cell.  
This state corresponds to the highest local density of states near the Fermi energy 
on these edge atoms. 
We can see that the wavefunction does not connect to either electrode and is 
localized to the lower layer. 
Sampling other wavefunctions along the $k$-path, we see that the states 
are likewise localized.

\begin{figure}[htb!]
\centering
\includegraphics[width=0.48\textwidth]{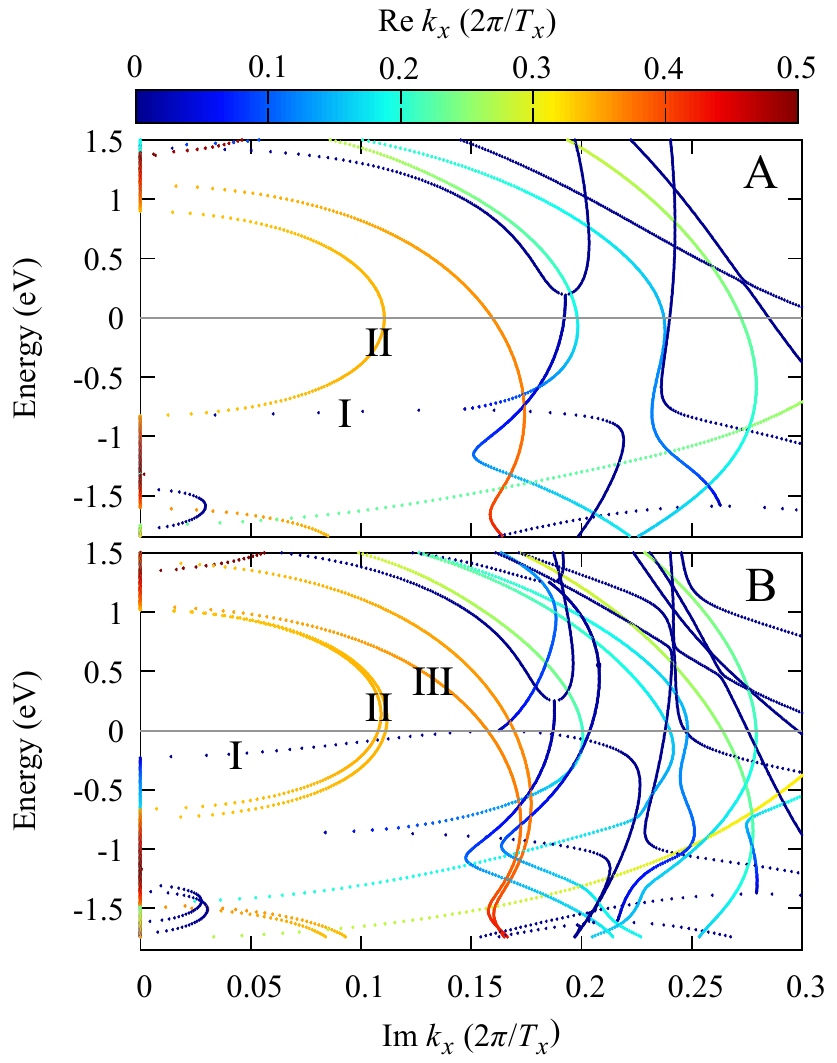}
\caption{\label{fig:cbands}
Complex band structure for (A) the monolayer and (B) the bilayer \ce{MoS_2} as in the planar and the interlayer junctions respectively. 
$k_y$ equals zero for both panels (A) and (B). 
The Fermi level is set to zero and marked by a gray solid line.}
\end{figure} 
In order to understand the in-plane decay of the electron wave function into the intrinsic region of the two \textit{p-i-n} junctions, we calculated the complex band structure of both monolayer and bilayer \ce{MoS2}. 
Fig. \ref{fig:cbands}A shows the complex band structure of monolayer \ce{MoS_2} to exhibit the evanescent states that decay in the transport direction of the planar junction. 
The reader may recall that the planar junction is in the $x$-$y$ plane and electron transport is along the $x$-direction. 
 The wavevector $k_y$ for the periodic $y$ direction is set to zero here since the electron transmission at this point is highest at the Fermi energy. 
There are eight species of evanescent states at Fermi energy for $\Im k_x < 0.3\, (2\pi/T_x) $, in which $T_x$ is the period of monolayer \ce{MoS2} in the transport direction of the planar junction. 
Two of them consist of slowly decaying states near the top of the valence band or the bottom of the conduction band. 
They are marked as species \rom{1} and \rom{2} in Fig.~\ref{fig:cbands}A. 
Hereafter, a valence top refers to a local maximum of the valence band under the constraint that $k_y$ is zero,  
and similarly for a conduction bottom. 
This is for the convenience of describing the connectivity between complex bands and real bands. 
Species \rom{1} derives from a valence top, and it is almost flat before $\Im k_x \sim 0.15\, (2\pi/T_x) $. 
The band flatness indicates that this species quickly becomes quickly decaying states as the energy increases from the valence top. 
After $\Im k_x \sim 0.15\,( 2\pi/T_x) $, species \rom{1} splits into two, an upper part and a lower part, with the upper part connecting to an unoccupied band much higher above the conduction bottom. 
Species \rom{2} connects a valence top and a conduction bottom, enclosing an area over the band gap. 
This area is minimal among all areas that are enclosed by complex bands over the forbidden gap. 
Within the Wentzel–Kramers–Brillouin approximation, the band-to-band tunneling probability decays exponentially with such a minimum area.~\cite{SLLRN285, SLLRN286}

Fig.~\ref{fig:cbands}B shows the complex band structure of a bilayer \ce{MoS_2}. 
In this case, there are three species of complex band which consist of slowly decaying states around the valence top or the conduction bottom. 
They are marked again by species \rom{1}, \rom{2}, and \rom{3} in Fig.~\ref{fig:cbands}B. 
Species \rom{1} of bilayer \ce{MoS_2} is similar to species \rom{1} of monolayer \ce{MoS_2} in three aspects: 
1) both of them are derived from the top of a valence band; 
2) a single band splits into two bands at $\Im k_x \sim 0.15\, (2\pi/T_x) $; and 
3) $\Re k_x$ equals zero before the band splitting. 
Note that $T_x$ of bilayer \ce{MoS2} is the same as that of monolayer \ce{MoS2}. 
Species \rom{2} of bilayer \ce{MoS_2} contains two copies of the dome-like complex band, each of which resembles species \rom{2} of monolayer \ce{MoS2}.
Species \rom{3} of bilayer \ce{MoS_2} is derived from the conduction bottom, and it extends into energies much lower than the valence top. 
For bilayer \ce{MoS2}, the minimum area of complex bands over the band gap is enclosed by both species \rom{1} and \rom{2}.
When species \rom{1} crosses species \rom{2}, an electron may transit between the two species with the aid of phonons since the two species are different in crystal momentum $\Re k_x$ and momentum is conserved during the transition. 
Phonon-assisted scattering may reduce the electron transmission and, thus, degrades the performance of the relevant electronic device.~\cite{SLLRN287} 

\onecolumngrid
\vskip 1\baselineskip

\begin{figure}[htb!]
\centering
  \includegraphics[width=0.85\textwidth,height=\textheight, keepaspectratio]{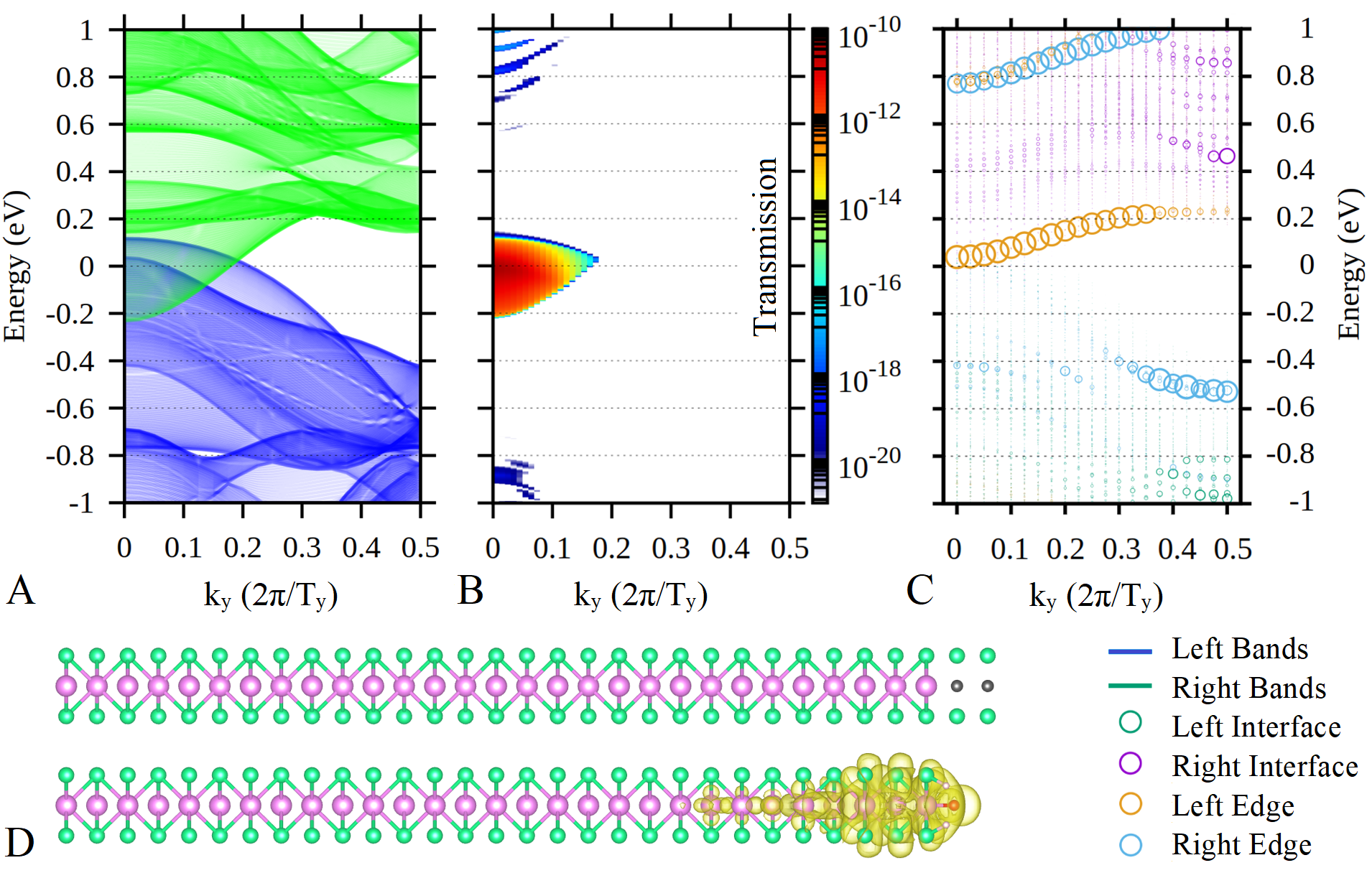}
\caption{A: The valence band of the left electrode and the conduction band of the right electrode. B: The transmission function of the interlayer junction at zero bias.  A maximum transmission of $ 3.6 \times 10^{-11}$ occurs at the gamma point at $ -7.5 \meV $.  The overlap of the electrode bands coincide with the transmission of the junction. C: Fat bands plots of contribution to the local density of states of pristine \ce{MoS_2} unit cells: the fat bands of unit cells touching edge-terminating atoms and the fat bands of unit cells touching doped atoms. D: The wavefunction corresponding to high LDOS near the Fermi energy seen in fig 5.c.  The state is localized to the edge atoms.
\label{Figure 4.}
}
\end{figure}

\twocolumngrid

\section{Conclusion}

We simulated armchair monolayer and interlayer \ce{MoS_2} $p$-$i$-$n$ junctions 
consisting of doped \ce{MoS_2} leads.  
Simulations were performed using $\mathrm{NEGF} + \mathrm{DFT}$ in order to
account for non-equilibrium conditions in the leads.  
Both junctions showed an initial positive differential resistance followed by a 
transition to negative differential resistance past a current maximum.  
Analysis of the partial density of states at different biases revealed a band alignment consistent with Esaki diode behavior.  
Furthermore, the band structures of both leads intersected at the same energy and 
$k$-points comprising the transmission of the interlayer junction.
These results confirm band to band tunneling as the primary mechanism of 
charge transfer in the low bias regime. 
Additionally, in-plane and interlayer band to band tunneling yields the 
same IV curve character, differing only in magnitude.
However, the two junctions significantly differ in Hartree potential, the interlayer junction having a large tunneling barrier in the van der Waals gap. The planar junction, being continuous in the transport direction, has no such electrostatic barrier.
Our complex band analysis reveals the slowest decaying evanescent states that contribute the most to the band-to-band tunneling probability. 
Phonon-assisted electron tunneling is more likely to occur in the interlayer junction than in the planar junction, degrading the performance of electronic devices based on bilayer \ce{MoS2}. 

Due to the known issues with edge termination in \ce{MoS_2} simulations, 
we used a H and O edge terminating scheme. 
The local density of states, obtained via fatband calculations, 
shows edge states near the Fermi energy. 
These states, however, are shown to have no effect on transport properties, 
as they are wholly localized to the edges of the interlayer junction.

\begin{acknowledgments}

This work was supported by the US Department of Energy (DOE), 
Office of Basic Energy Sciences (BES), under Contract No. DE-FG02-02ER45995. 
Computations were done using the utilities of National Energy Research Scientific Computing Center 
and University of Florida Research Computing.

\end{acknowledgments}



\bibliography{ref}

\end{document}